\documentclass[manuscript]{emulateapj}

\usepackage{amssymb}
\usepackage{amsmath}
\usepackage{booktabs}
\usepackage{multirow}
\usepackage{times}

\slugcomment{Accepted for publication in The Astrophysical Journal}

\shorttitle{The broad-band X-ray spectrum of PG\,1211+143}
\shortauthors{Lobban et al.}

\begin{document}

\title{Testing relativistic reflection and resolving outflows in PG\,1211+143 with {\it XMM-Newton} and {\it NuSTAR}}

\author{A. P. Lobban, K. Pounds and S. Vaughan}
\affil{University of Leicester, Department of Physics and Astronomy, University Road, Leicester, LE1 7RH, U.K.}
\email{al394@le.ac.uk}

\and

\author{J. N. Reeves}
\affil{Astrophysics Group, Keele University, Staffordshire, ST5 5BG, U.K. \\ Center for Space Science and Technology, 1000 Hilltop Circle, University of Maryland Baltimore County, Baltimore, MD 21250, USA}

\begin{abstract}

We analyze the broad-band X-ray spectrum (0.3--50\,keV) of the luminous Seyfert 1 / quasar PG\,1211+143 - the archetypal source for high-velocity X-ray outflows - using near-simultaneous {\it XMM-Newton} and {\it NuSTAR} observations.  We compare pure relativistic reflection models with a model including the strong imprint of photoionized emission and absorption from a high-velocity wind \citep{Pounds16a, Pounds16b}, finding a spectral fit that extrapolates well over the higher photon energies covered by {\it NuSTAR}.  Inclusion of the high $S/N$ {\it XMM-Newton} spectrum provides much tighter constraints on the model parameters, with a much harder photon index / lower reflection fraction compared to that from the {\it NuSTAR} data alone.  We show that pure relativistic reflection models are not able to account for the spectral complexity of PG\,1211+143 and that wind absorption models are strongly required to match the data in both the soft X-ray and Fe\,K spectral regions.  In confirming the significance of previously reported ionized absorption features, the new analysis provides a further demonstration of the power of combining the high throughput and resolution of long-look {\it XMM-Newton} observations with the unprecedented spectral coverage of {\it NuSTAR}.

\end{abstract}


\section{Introduction}

The standard model of an active galactic nucleus (AGN) is driven by accretion onto a supermassive black hole (SMBH; $M_{\rm BH} \sim 10^{6-9}$\,$M_{\odot}$).  AGN are powerful sources of X-rays which most likely originate close to the SMBH itself.  The X-ray spectra of unobscured AGN display an array of spectral features.  In particular, they are often dominated by a hard power-law component, thought to be produced when ultraviolet photons, emitted from an optically thick, geometrically thin accretion disc \citep{ShakuraSunyaev73}, are inverse-Compton scattered by a `corona' of hot electrons \citep{HaardtMaraschi93}.  Additional spectral features typically include a soft excess $< 2$\,keV \citep{ScottStewartMateos12}, a `Compton reflection' component $> 10$\,keV \citep{NandraPounds94} and emission lines, the strongest of which is often Fe\,K$\alpha$ fluorescence at $\sim$6.4\,keV \citep{GeorgeFabian91}.

Through systematic spectral studies with {\it ASCA}, {\it XMM-Newton}, {\it Chandra} and {\it Suzaku}, a significant fraction of AGN are now routinely observed to also exhibit strong signatures of ionized absorption in their X-ray spectra, with at least half of all AGN hosting photoionized ``warm'' absorbers (e.g. \citealt{ReynoldsFabian95, Blustin05}).  These absorbers produce numerous narrow and blueshifted absorption features (e.g. \citealt{Kaspi02, CrenshawKraemerGeorge03, McKernanYaqoobReynolds07}), implying material outflowing at a velocity from several hundred to several thousand km\,s$^{-1}$.

Through the study of blueshifted absorption lines of K-shell Fe, observations of higher-luminosity AGN have also revealed the presence of much more highly ionized absorbers originating in high-velocity disc winds ($v_{\rm out} \sim 0.1c$; e.g. PG\,1211+143; \citealt{Pounds03}, PDS 456; \citealt{ReevesOBrienWard03}).  Subsequent studies of archival {\it XMM-Newton} and {\it Suzaku} data have shown that similar high-velocity, highly-ionized outflows may be relatively common in nearby, luminous AGN (e.g. \citealt{Tombesi10, Tombesi11, Gofford13}).  In general, the derived outflow rates are comparable to the AGN accretion rate (up to several solar masses per year) and carry kinetic power on the order of a few per cent of the bolometric luminosity.  Such high-velocity outflows are believed to play a key role in linking growth of the SMBH and the host galaxy \citep{King03, King10} and offer an appealing explanation for the observed $M$--$\sigma$ relation for galaxies (Ferrarese \& Merritt 2000; Gebhardt et al. 2000).

The archetypal high-velocity-outflow source is PG\,1211+143, a luminous narrow-line Seyfert galaxy / quasar at a redshift of $z = 0.0809$ \citep{Marziani96}, which is both optically bright with a strong ``Big Blue Bump" and X-ray bright with a typical X-ray luminosity of $\sim$10$^{44}$\,erg\,s$^{-1}$.  The source is well-known for its spectral complexity and, through an {\it XMM-Newton} observation in 2001, provided the first detection of a sub-relativistic outflow ($v_{\rm out} \sim 0.14c$) in a non-BAL (broad absorption line) AGN \citep{Pounds03, PoundsPage06}.

Subsequent {\it XMM-Newton} observations of PG\,1211+143 were combined to model the photoionized absorption and emission spectra, quantifying the mass flux and energetics of the outflow and confirming the potential importance for galaxy feedback \citep{PoundsReeves07, PoundsReeves09, Pounds14}.  An extended {\it XMM-Newton} observation in 2014 has since shown the highly-ionized wind to have several outflow components, with primary velocities of $v_{\rm out} \sim 0.066c$ and $\sim$$0.129c$ \citep{Pounds16a}.

Surprisingly, the analysis of a 2014 {\it NuSTAR} observation of PG\,1211+143 \citep{Zoghbi15} found no evidence for the high-velocity outflow, leading to the suggestion that it may be a variable, transient feature - a result which would have important implications for the prevalence of high-velocity outflows in high-luminosity AGN and their wider significance for feedback.  Here, we report on a combined analysis of the same {\it NuSTAR} data with the contemporaneous extended $\sim$630\,ks  {\it XMM-Newton} observation obtained in 2014.

\section{Observations \& Data Reduction}

PG\,1211+143 was observed by {\it NuSTAR} \citep{Harrison13} four times in 2014: 2014-02-18 (ID: 60001100002), 2014-04-08 (60001100004), 2014-04-09 (60001100005) and 2014-07-07 (60001100007; hereafter {\it Nu}7), with exposure times of 111, 48, 64 and 74\,ks, respectively.  We used the \textsc{nupipeline} and \textsc{nuproducts} scripts, as part of \textsc{heasoft} v6.17\footnote{\url{https://heasarc.gsfc.nasa.gov/lheasoft/}}, to extract spectral products using the latest calibration database (version 20151008).  We extracted spectral products from circular source regions with radii of 52\,arc sec while background products were extracted from two same-sized circular regions separate from the source and away from the edges of the CCD.  We analyzed spectra from both of {\it NuSTAR}'s Focal Plane Modules (FPMA \& FPMB) using \textsc{xspec} v12.9.0 \citep{Arnaud96}.  We find that the {\it NuSTAR} spectra become heavily background-dominated $> 30$\,keV, resulting in the high-energy data becoming very noisy.  As such, we binned each individual spectrum in terms of signal-to-noise ($S/N$), ensuring that each spectral bin is detected at the $\geq 3\sigma$ level, which also allows us to use $\chi^{2}$ minimization. We analyzed all FPMA and FPMB spectra simultaneously, allowing for a floating cross-normalization parameter.  Typically, the cross-calibration normalization of the FPMB relative to the FPMA is $1.01 \pm 0.03$, consistent with the cross-calibration results presented in \citet{Madsen15}.  The FPMA+FPMB background-subtracted 3--50\,keV count rates of the four {\it NuSTAR} observations are $\sim$0.14, $\sim$0.19, $\sim$0.21 and $\sim$0.14\,ct\,s$^{-1}$, respectively.  This corresponds to respective 3--50\,keV observed fluxes of $\sim$6.4, $\sim$8.4, $\sim$9.2 and $\sim$6.6 $\times 10^{-12}$\,erg\,cm$^{-2}$\,s$^{-1}$.

\begin{figure}
\includegraphics[angle=0,scale=1.0]{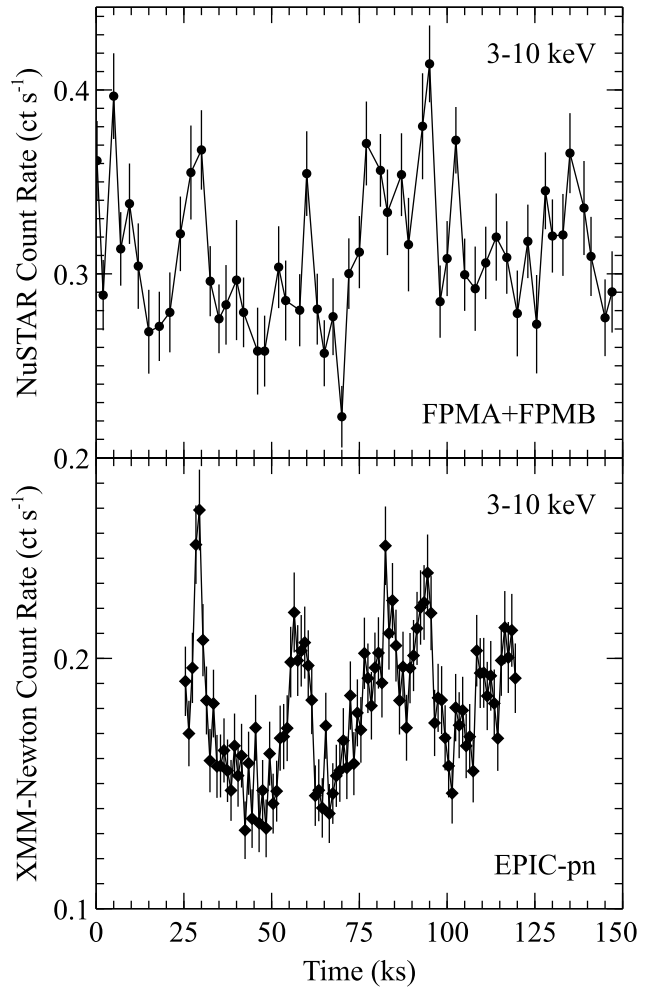}
\caption{Upper panel: the {\it Nu}7 FPMA+FPMB {\it NuSTAR} lightcurve from 3--10\,keV in 2\,ks bins.  The plotted count rate has been corrected using the \textsc{nulccorr} task.  Lower panel: the simultaneous rev2670 {\it XMM-Newton} EPIC-pn lightcurve from 3--10\,keV in 1\,ks bins.}
\label{fig:rev2670_nu7_lc}
\end{figure}

PG\,1211+143 was also observed with {\it XMM-Newton} \citep{Jansen01} on seven occasions in 2014 from 2014-06-02 to 2014-07-07, with a total exposure of $\sim$630\,ks.  The details of the observations and their subsequent data reduction is described in \citet{Lobban16}.  We note that the {\it XMM-Newton} spectra are binned up such that there are $> 100$\,ct\,bin$^{-1}$.  Here, we primarily focus on the fourth {\it NuSTAR} observation (referred to as {\it Nu7}), as this overlapped with the {\it XMM-Newton} campaign and was directly simultaneous with the final {\it XMM-Newton} observation (rev2670).  The four separate {\it NuSTAR} lightcurves are shown in \citet{Zoghbi15}.  The upper panel of Figure~\ref{fig:rev2670_nu7_lc} shows the {\it Nu}7 lightcurve\footnote{Note that the {\it NuSTAR} count rate has been adjusted via various corrections as part of the \textsc{nulccorr} task.  These include corrections for the `live time', point-spread function, vignetting, etc.}, while the lower panel shows the rev2670 {\it XMM-Newton} European Photon Imaging Camera (EPIC)-pn lightcurve in the 3--10\,keV energy band, demonstrating how the hard X-ray variability matches up.  However, given the lack of significant within-observation spectral variability at high energies, we proceed to analyze the time-averaged spectra.

\section{Spectral Analysis} \label{sec:spectral_analysis}

In \citet{Pounds16a,Pounds16b}, we reported the analysis of an extended {\it XMM-Newton} observation of PG\,1211+143, modelling both hard (2--10\,keV) spectra from the EPIC-pn camera \citep{Struder01} and higher-resolution soft ($< 2$\,keV) spectra from the Reflection Grating Spectrometer (RGS; \citealt{denHerder01}) with \textsc{xstar} \citep{Kallman96}.  We generated a set of self-consistent custom-built \textsc{xstar} grids with the correct ionizing continuum for PG\,1211+143, based on a spectral energy distribution model described in \citet{Lobban16}.  The grids were based on the abundances of \citet{GrevesseSauval98}.  We identified several components of ionized outflowing material, both confirming the presence of the high-velocity, sub-relativistic component ($\sim$0.14$c$) first detected in \citet{Pounds03} and detecting additional lower velocity components.  The continuum was modelled by a hard power law ($\Gamma \sim$ 1.7--1.8) with a black-body component and softer `unabsorbed' power law ($\Gamma \sim 3$) dominating $< 1$\,keV.  The softer power-law component is largely responsible for the long-term spectral variability observed on timescales of $\sim$days (see \citealt{Lobban16}).  We return to this model in Section~\ref{sec:broad-band_model}.

\begin{figure}
\includegraphics[angle=0,scale=1.0]{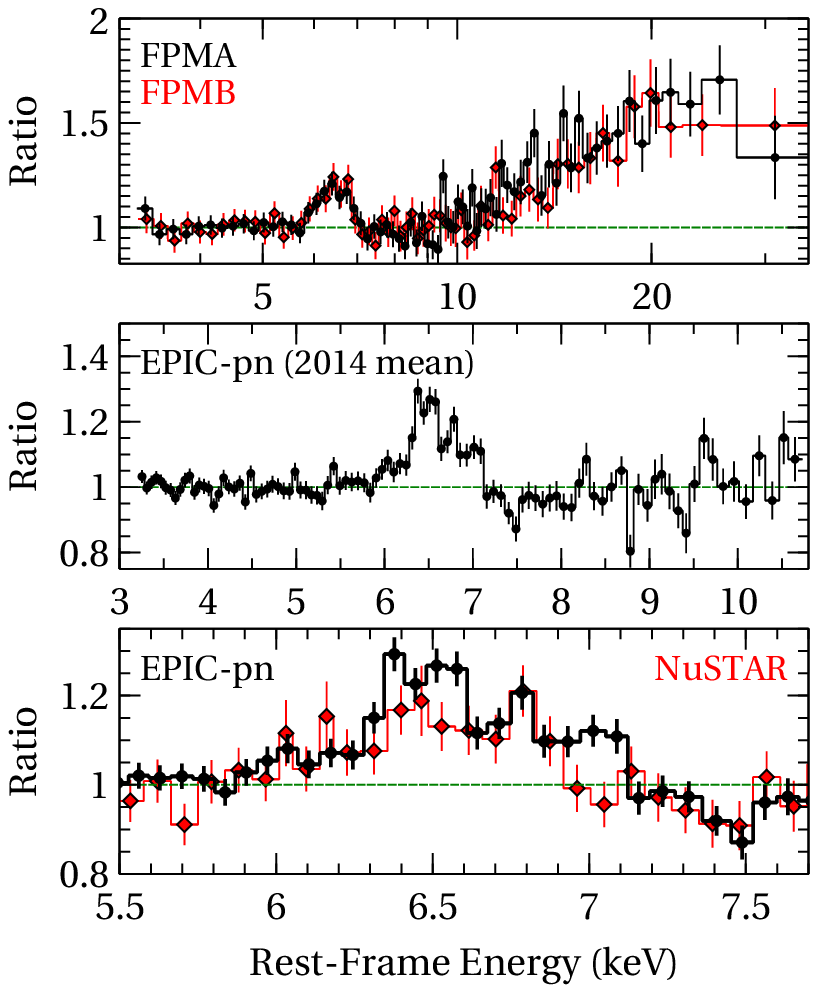}
\caption{The residuals of the co-added {\it NuSTAR} (upper panel) and {\it XMM-Newton} EPIC-pn (middle panel) spectra as a ratio to an absorbed power law fitted from 3--5 + 7.5--10\,keV.  The lower panel shows a zoom-in of the Fe\,K region in both the EPIC-pn (black) and combined {\it NuSTAR} (red) spectra.}
\label{fig:nu7_2014_po_ratio}
\end{figure}

Through the acquisition of contemporaneous {\it NuSTAR} observations, \citet{Zoghbi15} independently modelled the hard 3--50\,keV spectrum of PG\,1211+143.  They combined the data from both the FPMA and FPMB from all four observations in 2014.  Their best-fitting model comprised a single, steep power law ($\Gamma = 2.5 \pm 0.2$) plus strong emission from relativistic reflection, which accounts for the component of Fe\,K emission at $\sim$6--7\,keV and an excess of hard emission $> 10$\,keV, modelled using \textsc{relxill}\footnote{\url{http://www.sternwarte.uni-erlangen.de/~dauser/research/relxill/}} \citep{Dauser13}.  The reflection emission is strong, with $R = $2.5--4.5\footnote{The reflection fraction, $R$, is typically thought of as a proxy for the strength of the reflection component, where a value $R = 1$ corresponds to `standard' reflection from a semi-infinite disc covering $2\pi$\,sr, assuming a static corona.}, but the Fe abundance is sub-solar ($A_{\rm Fe} = 0.7 \pm 0.1$) to compensate for the weak Fe\,K line (relative to the reflection continuum).  And, highly relevant to our current study, they found no requirement for any signatures of high-velocity absorption.  In Figure~\ref{fig:nu7_2014_po_ratio}, we show the {\it NuSTAR} and {\it XMM-Newton} EPIC-pn residuals to a simple power-law fit.  The middle/lower panels highlight the structure in the Fe\,K band in the EPIC-pn spectrum, which spectral modelling finds to be consistent with contributions from narrow near-neutral Fe\,K$\alpha$, Fe\,\textsc{xxv} $1s$--$2p$ resonance and Fe\,\textsc{xxvi}\,Ly$\alpha$ emission lines at 6.4, 6.7 and 6.97\,keV, respectively.  Additionally, absorption structure is visible at $\sim$7.5\,keV.  In contrast, the {\it NuSTAR} spectra do not resolve this contribution from narrower absorption/emission lines with a smoother profile being observed.  This may be due to the difference in resolution between {\it XMM-Newton} and {\it NuSTAR} at Fe\,K, although the apparent lack of Fe\,\textsc{xxvi} emission at $\sim$6.97\,keV in the {\it NuSTAR} spectra may also arise from the effects of variability, given that the co-added {\it NuSTAR} observations span a time period of approximately five months.

All our subsequent fits include a Galactic column of $N^{\rm Gal}_{\rm H} = 3.06 \times 10^{20}$\,cm$^{-2}$ \citep{Willingale13}, modelled with the \textsc{tbabs} code within \textsc{xspec}, using the abundances of \citet{WilmsAllenMcCray00} and the absorption cross-sections of \citet{Verner96}.  In our broad-band EPIC-pn fits we ignore the 1.7--2.35\,keV band (observed frame; rest-frame $\sim$ 1.8--2.5\,keV) due to calibration uncertainties around the Si and Au detector edges at $\sim$1.8 and $\sim$2.2\,keV, respectively.  All errors are quoted at the 90\,per cent level for one parameter of interest (i.e. $\Delta \chi^{2} = 2.71$), unless stated otherwise.

\subsection{Fitting the hard X-ray spectrum with a relativistic reflection model} \label{sec:hard_band_relxill}

In this section, we detail the results of fitting the hard X-ray spectrum with \textsc{relxill}, a relativistic reflection model.  We use an up-to-date version of \textsc{relxill}, namely v0.4a (dated 18 January 2016), with the newly implemented reflection fraction parameter, $R$.  Previous versions of \textsc{relxill} used an alternative definition of $R$, as discussed in \citet{Dauser16}.  The interesting parameters in the \textsc{relxill} model are the photon index, $\Gamma$, reflection scaling fraction, $R$ (defined above), abundance of Fe relative to solar values, $A_{\rm Fe}$ (assuming the abundances of \citealt{GrevesseSauval98}), emissivity index, $q$, dimensionless black hole spin, $a^{*}$, inclination angle of the inner disc, $\theta$, inner and outer radii of emission, $r_{\rm in}$ and $r_{\rm out}$, respectively, ionization parameter\footnote{The ionization parameter is defined as $\xi = L_{\rm ion} / nr^{2}$, where $L_{\rm ion}$ is the ionizing luminosity from 1--1\,000\,Rydberg, $n$ is the gas density in units of cm$^{-3}$ and $r$ is the distance in units of cm. The parameter has units of erg\,cm\,s$^{-1}$.}, $\xi$, and high-energy cutoff, $E_{\rm cut}$, in units of keV.  Note that $r_{\rm in}$ and $r_{\rm out}$ can be computed in units of $r_{\rm g}$ (where $r_{\rm g} = GM/c^{2}$) or in units of the innermost stable circular orbit (ISCO; which is dependent on $a^{*}$).  In all fits, we fixed $r_{\rm out}$ at the default value 400\,$r_{\rm g}$, as the fits are insensitive to the parameter, and, as per \citet{Zoghbi15}, we assume that $q$ takes the form of a single power law.

\begin{table*}
\centering
\begin{tabular}{l c c c c c c c c c c}
\toprule
\multirow{2}{*}{Dataset (3--50\,keV)} & $\Gamma$ & $R$ & $A_{\rm Fe}$ & $q$ & $\theta$ & log\,$\xi$ & $r_{\rm in}$ & $a^{*}$ & $E_{\rm cut}$ & \multirow{2}{*}{$\chi^{2}_{\rm d.o.f.}$} \\  
& & & & & ($^{\circ}$) & & ($r_{\rm g}$) & & (keV) & \\
\midrule
Co-added {\it NuSTAR} & $2.35^{+0.16}_{-0.15}$ & $2.79^{+1.37}_{-0.91}$ & $0.86^{+0.83}_{-0.24}$ & $2.1^{+4.8}_{-1.3}$ & $36^{+13}_{-6}$ & $1.12^{+0.38}_{-0.98}$ & $<20$ & --- & $130^{+130}_{-36}$ & 563/520 \\
rev2670+{\it Nu}7 & $2.01^{+0.18}_{-0.08}$ & $1.16^{+1.11}_{-0.40}$ & $2.04^{+1.98}_{-1.05}$ & $1.9^{+6.1}_{-1.1}$ & $36^{+12}_{-6}$ & $1.95^{+0.31}_{-1.85}$ & $<90$ & --- & $>110$ & 673/662 \\
2014-mean-pn + {\it NuSTAR} & $1.95^{+0.07}_{-0.05}$ & $1.05^{+0.47}_{-0.49}$ & $1.70^{+0.77}_{-0.77}$ & $1.6^{+0.4}_{-0.7}$ & $40^{+7}_{-7}$ & $1.10^{+1.39}_{-0.76}$ & $<29$ & --- & $75^{+45}_{-24}$ & 1\,510/1\,237 \\
\bottomrule
\end{tabular}
\caption{The best-fitting values from our 3--50\,keV fits with \textsc{relxill}.  A dash (`---') means that we are unable to obtain useful constraints on that parameter.  The 90\,per cent uncertainties on the parameters are estimated from our MCMC output.}
\label{tab:relxill_parameters}
\end{table*}

We firstly fitted the co-added\footnote{The co-added spectra were created with the \textsc{addspec} code, as part of the \textsc{heasoft} suite, and were grouped in terms of $S/N$, such that each bin is detected at the $\geq 5\sigma$ level.} FPMA and FPMB {\it NuSTAR} spectra with \textsc{relxill} (where \textsc{relxill} models both an absorbed power law and the relativistic reflection component), and obtained parameters consistent with those of \citet{Zoghbi15} and a fit statistic of $\chi^{2}_{\rm d.o.f.} = 563/520$ (d.o.f. $=$ degrees of freedom).  We also find consistent results by fitting all four {\it NuSTAR} spectra simultaneously, allowing for a floating cross-normalization parameter to account for flux variability between observations ($\chi^{2}_{\rm d.o.f.} = 2\,089/2\,162$).  As per \citet{Zoghbi15}, we find that $q$ is degenerate with $r_{\rm in}$.  Additionally, we find no significant change in the fit between fixing $a^{*}$ at 0 (stationary) or 0.998 (maximal spin).  The {\it NuSTAR} spectra begin to become background-dominated $> 30$\,keV and the spectrum becomes very noisy - however, we fit up to 50\,keV here for comparison with \citet{Zoghbi15}.

We investigated the co-dependence of the parameters in this reflection model using a Markov Chain Monte Carlo (MCMC) method to explore the multi-dimensional parameter space.  Specifically, we used the \textsc{chain} command within \textsc{xspec}, using the Goodman-Weare algorithm and 200 walkers, where each chain contains a total of $10^{6}$ samples.  The output allows us to plot the regions of high likelihood for each parameter\footnote{If interpreted in Bayesian terms, we map the posterior density, implicitly assuming uniform priors on each fitted parameter.}.  Specifically, we used the \textsc{corner}\footnote{\url{https://zenodo.org/record/11020\#.Vs75hYyLR0t}} routine within the \textsc{python}\footnote{\url{https://www.python.org/}} programming language to create 1D histograms and 2D density plots, comparing all free parameters, and ensuring that our fits are not stuck in any local minima.

\begin{figure}
\includegraphics[angle=0,scale=1.0]{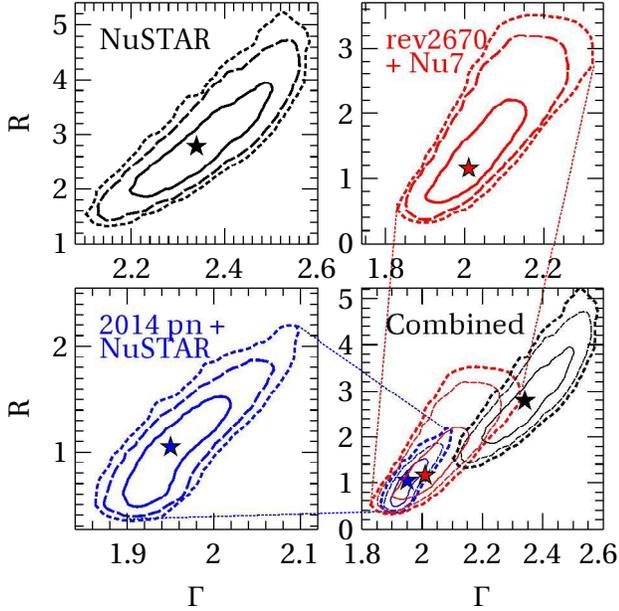}
\caption{2D density plots comparing $\Gamma$ and $R$ for the co-added {\it NuSTAR} spectrum (upper-left; black), the rev2670 EPIC-pn + {\it Nu}7 spectrum (upper-right; red) and the mean 2014 EPIC-pn + co-added {\it NuSTAR} spectrum (lower-left; blue) from a fit with \textsc{relxill} in the 3--50\,keV energy band.  The bottom-right panel shows the three contour plots superimposed on one another.  The projections were computed from a $10^{6}$-length MCMC run.  The inner to outer shaded contours contain 68.3, 95.5 and 99.7\,per cent of the total sample mass, respectively, and the stars represent the best-fitting values from the spectral fits.}
\label{fig:mcmc_contours}
\end{figure}

In Figure~\ref{fig:mcmc_contours}, we show a 2D density plot comparing $\Gamma$ and $R$ for the co-added {\it NuSTAR} spectrum (upper-left panel; black).  The three contour levels, from inner to outer, contain 68.3 (solid line), 95.5 (dashed line) and 99.7\,per cent (dotted line) of the total mass (i.e. the 1$\sigma$, 2$\sigma$ and 3$\sigma$ levels for a Gaussian distribution).  A steep $\Gamma$ / high $R$ solution is preferred: $\Gamma = 2.35^{+0.16}_{-0.15}$; $R = 2.79^{+1.37}_{-0.91}$, where the MCMC output is used to estimate the 90\,per cent uncertainties on our fit parameters.  Meanwhile, the fit prefers a sub-solar Fe abundance: $A_{\rm Fe} = 0.86^{+0.83}_{-0.24}$.  The best-fitting values are summarized in Table~\ref{tab:relxill_parameters}.

The upper-right and lower-left panels of Figure~\ref{fig:mcmc_contours} (red, blue) show the MCMC output when \textsc{relxill} is applied to: (i) the simultaneous rev2670 {\it XMM-Newton} EPIC-pn + {\it Nu}7 {\it NuSTAR} spectrum, and (ii) the total co-added 2014 EPIC-pn +  {\it NuSTAR} spectrum from 3--50\,keV.  The advantage of this approach is that the high resolution / throughput afforded by the EPIC-pn $< 10$\,keV allow for better constraints on the continuum and Fe\,K emission complex.  While case (i) provides us with a spectrum totally free from the uncertainties arising from long-term spectral variability, case (ii) affords us the highest $S/N$ in a PG\,1211+143 spectrum to date, with a combined $\sim$930\,ks of contemporaneous data.  Fitting the co-added spectrum is justified on the basis of the long-term X-ray variability being relatively weak $> 3$\,keV (see \citealt{Zoghbi15, Lobban16}).  All parameters were tied together between spectra except for the relative normalizations.  In the simultaneous fit [i.e. case (i)], the {\it XMM-Newton} and {\it NuSTAR} normalizations are found to be within $\pm 2$\,per cent of each other.

The inclusion of the EPIC-pn data places much tighter constraints on $\Gamma$ and $R$, with a harder $\Gamma$ / lower $R$ solution now preferred.  If we fix the photon index at $\Gamma = 2.51$, as per the value reported by \citet{Zoghbi15} from a fit to the {\it NuSTAR} data alone, we find that this is now rejected ($\Delta \chi^{2} = 35$), despite $R$ increasing to $\sim$5 and $A_{\rm Fe}$ dropping to $\sim$0.4 to compensate for the photon index being forced to a steeper value.  These tighter constraints help to demonstrate the power of combining the high throughput of the EPIC-pn camera with the broad-band coverage and unprecedented sensitivity $> 10$\,keV of {\it NuSTAR}.  

Finally, untying $\Gamma$ between datasets improves the fit in case (ii) ($\Delta \chi^{2} \sim 60$ with the value steepening in the {\it NuSTAR} spectrum by $\Delta \Gamma = 0.1$), most likely due to between-observation spectral variability, but not in case (i), with $\Gamma = 2.00 \pm 0.07$ for the {\it XMM-Newton} EPIC-pn and $\Gamma = 2.02 \pm 0.11$ for the simultaneous {\it NuSTAR} spectrum.  This difference of $\Delta \Gamma = 0.02$ lies within the cross-calibration uncertainties presented in \citet{Madsen15}.  Allowing $\Gamma$ to vary between datasets does not have any significant impact on $R$ in case (i).  In case (ii), $R$ is found to increase slightly ($\Delta R = 0.07$, with $A_{\rm Fe}$ dropping to $\sim$1 so as not to over-predict the Fe line), although this is still within the statistical uncertainties of the parameter and does not alter our conclusions.  None of the other parameters is significantly affected.  All free parameters and fit statistics are summarized in Table~\ref{tab:relxill_parameters}.

\subsection{Fitting the broad-band spectrum with relativistic reflection} \label{sec:broad-band_model_reflection}

\begin{figure}
\includegraphics[angle=0,scale=1.0]{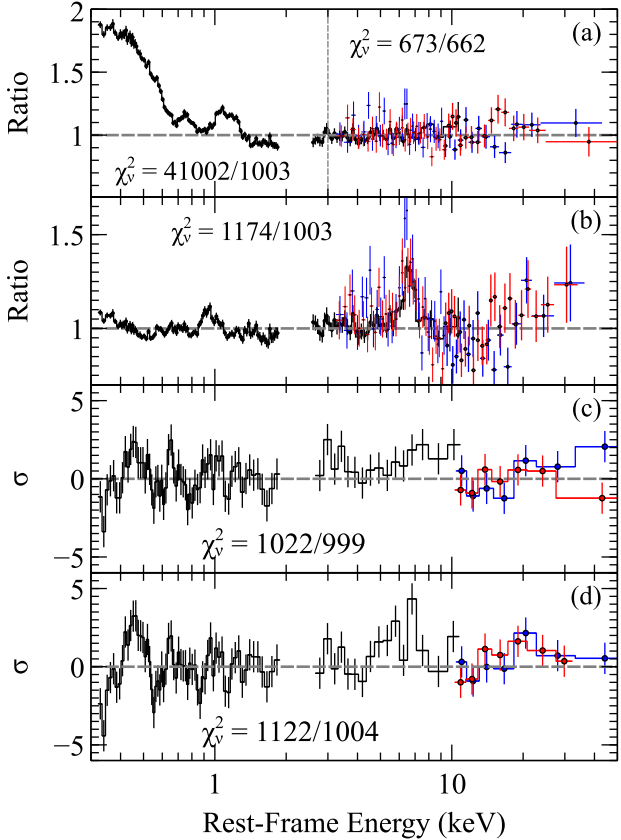}
\caption{Panel (a): an extrapolation of the \textsc{relxill} fit to low energies (rev2670 EPIC-pn = black; {\it Nu}7 FPMA = red; {\it Nu}7 FPMB = blue).  Panel (b): fitting the broad-band spectrum with \textsc{relxill}.  Panels (c) and (d) show the \textsc{relxill} + \textsc{xillver} + \textsc{pl}$_{\rm soft}$ vs \textsc{pl}$_{\rm hard}$ + \textsc{xillver} + \textsc{pl}$_{\rm soft}$ residuals, respectively.  The {\it NuSTAR} data are only plotted $> 10$\,keV in the lower panels.  This is for clarity as it allows the residuals of the two models to be better visually compared from 3--10\,keV where the FPMA and FPMB overlap with the EPIC-pn.}
\label{fig:relxill_residuals}
\end{figure}

While a relativistic reflection model provides a reasonable fit to the 3--50\,keV X-ray spectrum, in Figure~\ref{fig:relxill_residuals} we show the effect of extrapolating this model down to 0.3\,keV [panel (a)].  It is clear that the fit to the hard X-ray spectrum dramatically under-predicts the flux at low energies.  Here, we initially focus on the simultaneous rev2670 EPIC-pn + {\it Nu}7 spectrum.  Re-fitting the reflection model to the 0.3--50\,keV spectrum, the high $S/N$ soft X-ray spectral region drives the fit, returning $\Gamma$ and $R$ to higher values of $\sim$2.4 and $\sim$5, respectively, and pushing the emissivity up to $q \sim 5$ and $a^{*}$ to maximal spin with most of the emission arising from the ISCO.  This produces a strong, relatively smooth soft excess $< 2$\,keV, but such strong relativistic blurring blends the Fe\,K emission complex into the continuum.  This leaves significant residuals and results in a relatively poor fit: $\chi^{2}_{\rm d.o.f.} = 1\,174/1\,003$ [panel (b)] and $\chi^{2}_{\rm d.o.f.} = 239/127$ in the Fe\,K band (5.5--7.5\,keV).  

Panel (c) of Figure~\ref{fig:relxill_residuals} shows the $\sigma$-residuals of a dual-reflector fit, incorporating additional emission from a non-relativistic distant reflector using the \textsc{xillver} model of \citet{Garcia14}, tying $\Gamma$, $A_{\rm Fe}$, $\theta$ and $E_{\rm cut}$ to the corresponding \textsc{relxill} parameters.  The ionization parameter of the \textsc{xillver} component is consistent with zero as it matches fluorescent emission from Fe\,K$\alpha$ while modelling additional Compton reflection from optically-thick material.  The addition of this component improves the fit by $\Delta \chi^{2} = 55$, although the fit in the soft band is still poor.  As such, we also include an additional steep power law component ($\Delta \chi^{2} = 97$), accounting for a component of soft excess, as was required by the RGS analysis of \citet{Pounds16b}.  The steep component is the dominant source of long-term spectral variability (e.g. see the difference spectrum in \citealt{Lobban16}).  

Meanwhile, panel (d) of Figure~\ref{fig:relxill_residuals} shows the $\sigma$-residuals to a broad-band fit with no relativistic reflection (i.e. with the hard and soft power laws plus the distant, near-neutral reflector).  This helps to visualize which spectral components the relativistic reflector is modelling.  It appears that the residuals are very similar in the two cases, with an exception at the Fe\,K complex, where \textsc{relxill} models some excess emission from $\sim$5.5--7\,keV.  Some of this excess emission may arise from a broadened component of emission, although some of the residuals, particularly in the EPIC-pn spectrum, may be accounted for with additional narrow emission lines [i.e. see \citealt{Pounds16a} where the stacked EPIC-pn spectrum resolves the Fe\,K emission complex consisting of Fe\,K$\alpha$ fluorescence plus resonance lines from He-like ($\sim$6.7\,keV) and H-like ($\sim$6.97\,keV) Fe (see Figure~\ref{fig:nu7_2014_po_ratio}; lower panel), which are well modelled with photoionized absorption and may represent scattering of the wind].  Overall, the dual-reflector model gives an acceptable fit to the simultaneous broad-band spectrum: $\chi^{2}_{\rm d.o.f.} = 1\,022/999$.  The best-fitting parameters are detailed in Table~\ref{tab:dual-reflector_parameters}.

\begin{table}
\centering
\begin{tabular}{l c c c}
\toprule
Model & \multirow{2}{*}{Component} & rev2670 & 2014-mean-pn \\
(Statistic) & & + {\it Nu}7 & + {\it NuSTAR} \\
\midrule
\multirow{10}{*}{\textsc{relxill}} & $\Gamma$ & $2.21^{+0.05}_{-0.06}$ & $2.20^{+0.02}_{-0.02}$ \\
& $R$ & $4.48^{+0.54}_{-1.88}$ & $4.33^{+0.55}_{-0.40}$ \\
& $A_{\rm Fe}$ & $2.11^{+0.38}_{-0.42}$ & $1.35^{+0.22}_{-0.24}$ \\
& $q$ &$3.6^{+0.5}_{-0.7}$ & $4.0^{+0.5}_{-0.3}$ \\
& $\theta$ ($^{\circ}$) & $44^{+4}_{-6}$ & $44^{+2}_{-2}$ \\
& log\,$\xi$ & $1.00^{+0.07}_{-0.13}$ & $1.17^{+0.08}_{-0.06}$ \\
& $r_{\rm in}$ ($r_{\rm g}$) & $<3.3$ & $<2.0$ \\
& $a^{*}$ & $>0.69$ & $>0.97$ \\
& $E_{\rm cut}$ (keV) & $130^{+80}_{-60}$ & $>150$ \\
& log\,($N$) & $-4.81^{+0.04}_{-0.05}$ & $-4.73^{+0.03}_{-0.02}$ \\
\midrule
\multirow{7}{*}{\textsc{xillver}} & $\Gamma$ & $2.21^{\dagger}$ & $2.26^{\dagger}$ \\
& $A_{\rm Fe}$ & $2.11^{\dagger}$ & $1.35^{\dagger}$ \\
& $\theta$ ($^{\circ}$) & $44^{\dagger}$ & $44^{\dagger}$ \\
& log\,$\xi$ & $<0.13$ & $<0.03$ \\
& $E_{\rm cut}$ (keV) & $130^{\dagger}$ & $>150^{\dagger}$ \\
& log\,($N$) & $-5.42^{+0.15}_{-0.26}$ & $-5.36^{+0.14}_{-0.11}$ \\
& $\Delta \chi^{2}$ & 55 & 99 \\
\midrule
\multirow{3}{*}{\textsc{pl}$_{\rm soft}$} & $\Gamma$ & $3.56^{+0.24}_{-0.08}$ & $3.66^{+0.12}_{-0.11}$ \\
& log\,($N$) & $-3.41^{+0.10}_{-0.11}$ & $-3.59^{+0.11}_{-0.05}$ \\
& $\Delta \chi^{2}$ & 97 & 413 \\
\midrule
Fit statistic & $\chi^{2}_{\rm d.o.f.}$ & $1\,022/999$ & $3\,573/3\,302$ \\
\bottomrule
\end{tabular}
\caption{The best-fitting values from our broad-band 0.3--50\,keV fits with the dual-reflector model described in Section~\ref{sec:broad-band_model_reflection}.  Note that $\Gamma$, $A_{\rm Fe}$, $\theta$ and $E_{\rm cut}$ are tied between the \textsc{relxill} and \textsc{xillver} models (denoted by $^{\dagger}$), while `log\,($N$)' refers to the normalization of the model component, where the normalization is in units of photon\,cm$^{-2}$\,s$^{-1}$.  The quoted value of $\Gamma$ in the right-hand column was obtained from the EPIC-pn spectrum; the {\it NuSTAR} values are typically around $\Gamma \sim 2.3$.  The $\Delta \chi^{2}$ values refer to the improvement in the fit statistic upon sequentially adding the model components (i.e. \textsc{relxill} vs \textsc{relxill} $+$ \textsc{xillver} and \textsc{relxill} $+$ \textsc{xillver} vs \textsc{relxill} $+$ \textsc{xillver} $+$ \textsc{pl}$_{\rm soft}$, respectively).
}
\label{tab:dual-reflector_parameters}
\end{table}

\begin{figure}
\includegraphics[angle=0,scale=1.0]{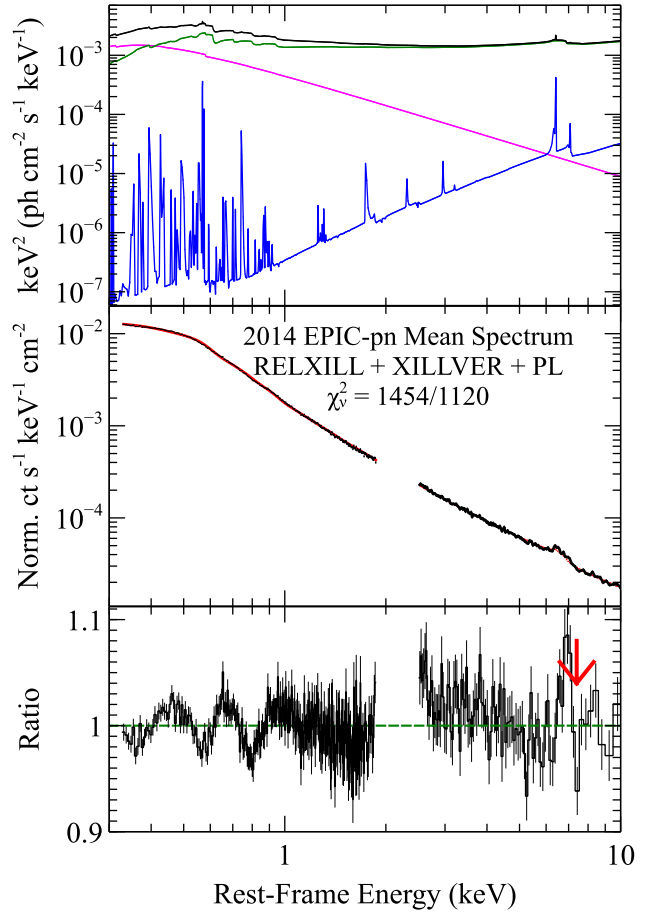}
\caption{Upper panel: the two-reflector model fitted to the mean EPIC-pn spectrum; the summed contribution is shown in black while the \textsc{relxill}, \textsc{xillver} and soft power-law components are shown in green, blue and magenta, respectively.  Middle panel: the mean EPIC-pn spectrum fitted with the two-reflector model (red).  Lower panel: the ratio of the residuals with strong signatures of ionized absorption at $\sim$0.5--0.9\,keV and $\sim$7.5\,keV (red arrow) and ionized emission at $\sim$6.7--6.9\,keV.}
\label{fig:2014_pn_relxill}
\end{figure}

We also apply the same model to the mean EPIC-pn $+$ {\it NuSTAR} spectrum, including all four {\it NuSTAR} observations separately, accounting for changes in flux over time with a floating cross-normalization parameter.  The final fit statistic is $\chi^{2}_{\rm d.o.f.} = 3\,573/3\,302$, where we have also allowed $\Gamma$ to vary between {\it NuSTAR} observations.  The best-fitting parameters are detailed in Table~\ref{tab:dual-reflector_parameters}.  Figure~\ref{fig:2014_pn_relxill} shows the effect of fitting the dual-reflector model to the mean EPIC-pn spectrum.  With the increased $S/N$ afforded by $\sim$630\,ks of data, strong residuals are now apparent in the spectrum ($\chi^{2}_{\rm d.o.f.} = 1\,454/1\,120$).  In particular, there is clear evidence of ionized absorption from $\sim$0.5--0.9\,keV plus an excess of positive residuals just blue-ward of the peak of the Fe\,K emission complex (i.e. $> 6.4$\,keV).  This is most likely the signature of the narrow He-like / H-like resonance lines from Fe reported in \citet{Pounds16a}.  Additionally, there are significant negative residuals at $\sim$7.5\,keV, which is a signature of the highest-velocity component of the fast outflow ($v \sim 0.13c$; see \citealt{Pounds16a}).  Indeed, the fit is very poor in these bands: 0.5--0.9\,keV: $\chi^{2}_{\rm d.o.f.} = 181/68$; 5.5-7.5\,keV: $\chi^{2}_{\rm d.o.f.} = 233/165$.  In summary of this section, we find that combining {\it XMM-Newton} and {\it NuSTAR} spectra shows a model based on reflection alone cannot reproduce all of the observed spectral features without including the now well-established ionized emission and absorption in both the Fe\,K and soft X-ray bands.

\subsection{An alternative fit to the broad-band spectrum with an ionized absorption model} \label{sec:broad-band_model}

In this section, we demonstrate an alternative approach with a model incorporating the ionized wind parameters from \citet{Pounds16a, Pounds16b} to form a broad-band model which we now fit to the combined {\it XMM-Newton}+{\it NuSTAR} spectrum.  As noted above, those two papers report on analyses of the EPIC-pn hard X-ray band and the soft RGS band, respectively.  In the former case, the 2--10\,keV baseline X-ray continuum is dominated by a hard power law ($\Gamma \sim 1.7$) with multiple blue-shifted absorption lines corresponding to two ionized outflow components, with parameters of log\,$\xi = 4.0$ and 3.4, $N_{\rm H} = 3.7$ and 2.0 $\times 10^{23}$\,cm$^{-2}$ and $v_{\rm out} = 0.129$ and $0.066c$, respectively, when modelled with \textsc{xstar} \citep{Kallman96}.  The near-neutral Fe\,K$\alpha$ fluorescence line at 6.4\,keV is modelled together with a modest reflection continuum using a distant reflector (\textsc{xillver}), whose photon index is tied to that of the hard power law.  Additional \textsc{xstar} emission grids model ionized emission lines at $\sim$6.7 and $\sim$6.97\,keV, identified with He-like and H-like Fe $1s$--$2p$ resonance transitions.

The baseline continuum for the RGS analysis (0.3--2\,keV) is dominated by the well-known `soft excess', with both ionized emission and absorption features being imprinted on a black-body component ($kT \sim 0.1$\,keV).  An additional soft power law ($\Gamma \sim 2.9$), found in inter-orbit difference spectra \citep{Lobban16}, matches the variable component of the soft excess and is included in the model.  The analysis by \citet{Pounds16b} finds evidence for several outflow components, ranging in log\,$\xi$ from 1.5--3.2 and $N_{\rm H}$ from 1.5--28 $\times 10^{20}$\,cm$^{-2}$, with lower-ionization/lower-column-density counterparts apparently co-moving with the dual-velocity outflow components detected in the hard X-ray band.  Finally, two zones of emission are included to model narrow photoionized emission lines, which indicate an average covering factor of $\sim$50\,per cent.

We now tie both hard- and soft-band models together to generate a single broad-band model, which, in \textsc{xspec}, takes the form: \textsc{tbabs} $\times$ [\textsc{pl}$_{\rm soft}$ + (\{\textsc{pl}$_{\rm hard}$ + \textsc{bbody}\} $\times$ \textsc{xstar}$^{\rm abs.}$) + \textsc{xstar}$^{\rm ems.}$ + \textsc{xillver}].  We firstly focus on the simultaneous rev2670 EPIC-pn + {\it Nu}7 spectrum to be sure that we are free from any complications arising from spectral variability between observations.  Due to the more modest spectral resolution $< 2$\,keV of the EPIC-pn compared to the RGS, we fix the best-fitting parameters of the ionized absorption/emission grids at the values quoted in \citet{Pounds16b}, just allowing the column densities to vary.  Fitting to the 0.3--10\,keV EPIC-pn data alone gives a good fit: $\chi^{2}_{\rm d.o.f.} = 475/471$, with $\Gamma_{\rm hard} \sim 1.8$ and $\Gamma_{\rm soft} \sim 3.0$.  We then include the simultaneous {\it NuSTAR} data and extend the fit to 50\,keV (although, as mentioned in Section~\ref{sec:hard_band_relxill}, the data become background-dominated and noisy $> 30$\,keV).  Without re-fitting or even re-normalizing, the change in the fit statistic is $\Delta \chi^{2} = 583$ for $523$ d.o.f. ($\chi^{2}_{\rm d.o.f.} = 1\,058/994$), indicating that our fit to the EPIC-pn data predicts the hard X-ray continuum $>10$\,keV well.  Allowing the model to re-fit (including a floating cross-normalization factor for each {\it NuSTAR} detector and allowing $\Gamma$ to vary slightly between the {\it XMM-Newton} and {\it NuSTAR} spectra) improves the fit to $\chi^{2}_{\rm d.o.f.} = 1\,003/990$.  The parameters of the Fe\,K emission / absorption zones are largely consistent with \citet{Pounds16a}, varying slightly in $N_{\rm H}$ and log\,$\xi$, while the reflector ({\textsc{xillver}) has an emission peak at $\sim$6.45\,keV (log\,$\xi \sim 0.9$; $A_{\rm Fe} \sim 2.5$).  The individual components of the broad-band model are shown in Figure~\ref{fig:broad-band} (upper panel) along with the residuals to the best fit (middle panel).  

\begin{figure}
\includegraphics[angle=0,scale=1.0]{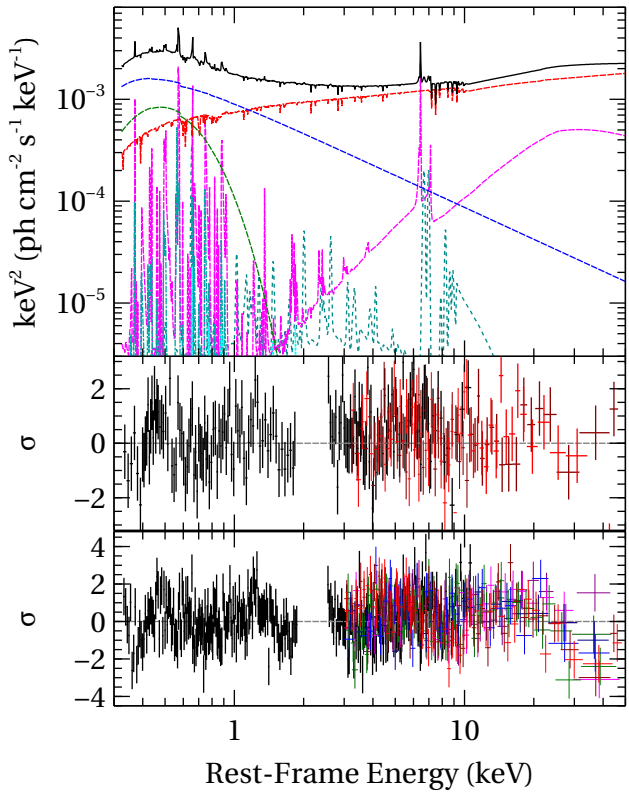}
\caption{Upper panel: the additive components contributing to the full broad-band model (black): $\Gamma_{\rm hard}$ (red), $\Gamma_{\rm soft}$ (blue), black body (green), \textsc{xillver} (magenta), ionized emission (cyan).  Middle panel: 1-$\sigma$ residuals of the rev2670  EPIC-pn (black) + {\it Nu}7 (red) 0.3--50\,keV spectrum.  Lower panel: 1-$\sigma$ residuals of the mean 2014 EPIC-pn (black) + Nu2 (red) + Nu4 (blue) + Nu5 (green) + {\it Nu}7 (magenta) spectrum.}
\label{fig:broad-band}
\end{figure}

We then fit the model to the high $S/N$ 2014 mean EPIC-pn $+$ {\it NuSTAR} spectrum.  We include all four {\it NuSTAR} observations separately with a floating cross-normalization parameter to account for changes in flux over time.  Extending the fit to 50\,keV results in $\chi^{2}_{\rm d.o.f.} = 3\,490/3\,377$, where we also allow $\Gamma_{\rm hard}$ to vary between observations (which typically falls within the range $\Gamma_{\rm hard} =$ 1.69--1.94).  Meanwhile, $\Gamma_{\rm soft}$ remains at a value of $\sim$3.1.  We again find that the best-fitting parameters of outflowing ionized absorption components are consistent with those reported in \citet{Pounds16a, Pounds16b}.  The residuals are shown in Figure~\ref{fig:broad-band} (lower panel).

Having accounted for the absorption due to both the high-velocity wind apparent in the Fe\,K band and the lower-velocity/ionization counterparts in the soft X-ray band, we now measure the strength of any relativistic reflection component allowed by the data.  We use the model described above but replace the primary power-law component (\textsc{pl}$_{\rm hard}$) with \textsc{relxill}, tying $\Gamma$, $A_{\rm Fe}$, $\theta$ and $E_{\rm cut}$ to the corresponding \textsc{xillver} values, like in Section~\ref{sec:broad-band_model_reflection}.

In the simultaneous rev2670 + {\it Nu}7 case, this results in a slight improvement to the fit with $\Delta \chi^{2} = 25$.  The best-fitting reflection fraction is $R \sim 1.1$, while the Fe abundance and photon index remain at values of $A_{\rm Fe} \sim 3$ and $\Gamma \sim 1.8$, respectively.  The \textsc{relxill} component is moderately ionized (log\,$\xi \sim 1.3$) and predominantly manifests itself by including some degree of broadening on the red side of the Fe\,K complex.  Meanwhile, the \textsc{xillver} component models the near-neutral component of Fe\,K$\alpha$ emission (log\,$\xi = 0$).  

A very similar result is reached by performing the same test on the 2014 mean EPIC-pn $+$ {\it NuSTAR} spectrum ($\Delta \chi^{2} = 52$), resulting in a modest reflection strength, $R \sim 0.6$, while the photon index and Fe abundance remain unchanged: $\Gamma \sim 1.8$; $A_{\rm Fe} \sim 3$.  The remaining key parameters in the \textsc{relxill} model are largely consistent with those reported in Table~\ref{tab:dual-reflector_parameters}; i.e. log\,$\xi \sim 1.2$, $q \sim 3$, $\theta \sim 40$, $r_{\rm in} < 6$.  The black hole spin, $a^{*}$, is unconstrained although we find a lower-limit on the high-energy cutoff of $E_{\rm cut} > 130$\,keV.

Finally, we return to check for the presence of highly-ionized absorption in the contemporaneous {\it NuSTAR} data.  In terms of the EPIC-pn analysis, the two absorption grids modelling the highly-ionized two-velocity outflow are highly significant, with a combined contribution to the fit statistic of $\Delta \chi^{2} = 42$ (for six additional parameters; meanwhile, the highly-ionized emission and soft-band absorption grids improve the fit by $\Delta \chi^{2} = 39$ and $\Delta \chi^{2} = 44$, respectively).  \citet{Zoghbi15} detected no significant discrete absorption features in the {\it NuSTAR} spectrum.  We now find, in direct comparison with {\it XMM-Newton}, including the same two \textsc{xstar} absorption zones to the same baseline model fitted to just the contemporaneous {\it NuSTAR} observations offers a far less significant improvement to the fit statistic, with $\Delta \chi^{2} = 9$.  As such, even though the highly-ionized outflow is now well established thanks to {\it XMM-Newton}, we also confirm the main result from \citet{Zoghbi15} that it was not detectable by {\it NuSTAR} alone.  This demonstrates the importance of long-look observations with {\it XMM-Newton} in addition to the broad-band coverage of {\it NuSTAR}.  

\section{Discussion \& Conclusions} \label{sec:discussion}

An extended {\it XMM-Newton} observation of PG\,1211+143 has revealed spectral structure in the Fe\,K band, unseen in previous (shorter) observations.  Utilizing the high throughput of the EPIC-pn, \citet{Pounds16a} showed that highly-ionized outflow resolved two velocity components ($v_{\rm out} = 0.129$ and $0.066c$), while an analysis of the high-resolution RGS data in \citet{Pounds16b} found lower-ionization co-moving counterparts of that dual-velocity flow.  Here, we have extended our modelling over a wider energy range, incorporating contemporaneous {\it NuSTAR} data.  The power of this approach -- i.e. combining {\it XMM-Newton} and {\it NuSTAR} data -- has previously been demonstrated in a precise measurement of the higher-energy continuum and thereby helping to disentangle absorption and reflection in a number of AGN - e.g. NGC 1365 \citep{Risaliti13, Walton14, Rivers15}, NGC 5548 \citep{Kaastra14, Ursini15, Cappi16} and PDS 456 \citep{Nardini15}.

In Section~\ref{sec:hard_band_relxill}, we fitted the hard X-ray spectrum with relativistic reflection, following the recent analysis of \citet{Zoghbi15}.  Using the {\it NuSTAR} data alone, we confirm their solution, finding a steep photon index ($\Gamma \sim 2.4$) and high reflection fraction ($R \sim$ 2--3).  However, by also including the simultaneous {\it XMM-Newton} spectrum, just in the 3--10\,keV range (which overlaps with the {\it NuSTAR} bandpass), a harder-$\Gamma$ / lower-$R$ solution is strongly preferred.  Replacing the overlapping {\it XMM-Newton} data with the high $S/N$ stacked 2014 EPIC-pn spectrum further tightens the constraints, with $\Gamma = 1.95^{+0.07}_{-0.05}$ and $R = 1.05^{+0.47}_{-0.49}$.  Such continuum parameters are now fully consistent with those reported by \citet{Pounds16a, Pounds16b}, where a detailed analysis of the long-look {\it XMM-Newton} observation is performed, taking into account the spectral imprints of the outflowing wind.

In Section~\ref{sec:broad-band_model_reflection}, we attempted to fit the broad-band (0.3--50\,keV) spectrum with relativistic reflection, finding that, while \textsc{relxill} alone cannot simultaneously fit the excess of emission in the soft band and the significant Fe\,K emission complex, a two-reflector model (comprising a distant reflector and a relativistically blurred reflector) provides a good fit to the simultaneous rev2670+{\it Nu}7 spectrum.  However, Figure~\ref{fig:2014_pn_relxill} illustrates the inability of this model to match the spectral features of the stacked 2014 EPIC-pn spectrum.  Clearly, a pure reflection model cannot match fine structure in the high $S/N$ spectrum without significant modification - in particular, strong signatures of ionized absorption at $\sim$0.5--0.9\,keV and $\sim$7.5\,keV remain in the residuals, along with an excess of ionized emission at $\sim$6.7--6.9\,keV - spectral features resolved by the excellent statistics of the {\it XMM-Newton} data and identified in \citet{Pounds16a, Pounds16b}.  This emphasises the value of long-look observations, where the $\sim$630\,ks stacked EPIC-pn spectrum reveals significant spectral structure, otherwise lost in the statistical noise of shorter observations.

In Section~\ref{sec:broad-band_model} we explore an alternative broad-band 0.3--50\,keV {\it XMM-Newton} + {\it NuSTAR} fit, now including the ionized absorption/emission components required by the {\it XMM-Newton} spectrum.  We find that the model presented in \citet{Pounds16a, Pounds16b} extrapolates well to the {\it NuSTAR} band and, overall, provides an excellent fit to the broad-band X-ray spectrum.  Having thus accounted for the various components of the high-velocity outflowing absorber, we note that a contribution of relativistic reflection is allowed by the data, manifesting itself in the form of excess emission just red-ward of the Fe\,K complex, although its strength is rather modest ($R \sim 0.6$).  Additionally, when fitted to the 0.3--50\,keV band, the absorption-dominated model requires a best-fitting photon index of $\Gamma \sim 1.8$, which is significantly different from the value of $\Gamma \sim 2.4$ arrived at by fitting the {\it NuSTAR} data alone.  This conclusion is, of course, supported by our MCMC analysis which shows that a harder-$\Gamma$/lower-$R$ solution is preferred when including the simultaneous {\it XMM-Newton} data, regardless of the assumed continuum model.

Finally, we comment on one of the headline results of the \citet{Zoghbi15} analysis, namely that the {\it NuSTAR} spectrum exhibits no evidence of the high-velocity outflow originally reported in \citet{Pounds03}.  This claim was based on a line-detection search across the appropriate bandpass with emphasis placed on the $\sim$7.1\,keV absorption line, first detected in the 2001 observation.  The non-detection of that absorption line in the {\it NuSTAR} spectrum led to the suggestion that the ultra-fast outflow in PG\,1211+143 may be transient, or at least highly variable, implying that the contribution of ultra-fast outflows to feedback could be substantially lower than previously thought.

\citet{Pounds16a} find that this absorption is indeed variable, depending on both column density and ionization parameter, and had a mean equivalent width $\sim$3 times smaller during the 2014 {\it XMM-Newton} observation than it was when first detected in 2001.  As such, it is unsurprising that the feature was not significantly detected by a single line search in the {\it NuSTAR} spectrum.  More generally, while a line search has provided a useful method in the past for detecting individual strong resonance lines, a preferred approach is to use self-consistent photoionization models (e.g. with \textsc{xstar}), as we do here, which incorporate a physically realistic array of common-velocity lines and absorption edges from various ions (including additional higher-order lines) and imprint appropriate related continuum curvature, for a given $N_{\rm H}$, $\xi$ and $v_{\rm out}$.  The improvement in the fit statistic is then computed from the summed contributions of all of the absorption features, where several individual weak features may have a significant impact overall (although we do note that even this approach would not have led to a significant detection of the outflow using the contemporaneous {\it NuSTAR} data alone).

In summary, while \citet{Zoghbi15} detected no discrete absorption features in the {\it NuSTAR} spectrum, the extended {\it XMM-Newton} observations, also in 2014, confirm that, not only is the outflow still present, but that it exhibits more complex spectral structure than previously realized.

In studying the properties of powerful AGN winds, now widely accepted as one of the most important outcomes of the past decade of X-ray observations, we have noted the importance of self-consistent modelling with physically-motivated absorption models which properly account for multiple ions and continuum curvature, and emphasise that (i) finding a common velocity for a line series can provide a highly significant result, and (ii) {\it XMM-Newton} remains a powerful observatory for detailed, high-resolution spectroscopy, particularly when used in tandem with the broad-band coverage of {\it NuSTAR}.

\acknowledgments

This research has made use of the NASA Astronomical Data System (ADS), the NASA Extragalactic Database (NED) and is based on observations obtained with {\it XMM-Newton}, an ESA science mission with instruments and contributions directly funded by ESA Member States and NASA, and the {\it NuSTAR} mission, a project led by the California Institute of Technology (Caltech), managed by the Jet Propulsion Laboratory (JPL).  APL and SV acknowledge support from STFC consolidated grant ST/K001000/1 and JNR acknowledges financial support via NASA grant NNX15AF12G.  We thank an anonymous referee for highly constructive comments which helped improve the clarity of the paper.

{\it Facilities:} \facility{{\it XMM-Newton} (EPIC)}, \facility{{\it NuSTAR}}.

\clearpage

\end{document}